\documentclass[12pt, preprint]{aastex}
\usepackage{graphicx}

\shorttitle{Masses and Radii for five binary stars }

\shortauthors{Fernandez et al.}

\newcommand\kms{\ifmmode{\rm km\thinspace s^{-1}}\else km\thinspace s$^{-1}$\fi}
\newcommand\ms{\ifmmode{\rm m\thinspace s^{-1}}\else m\thinspace s$^{-1}$\fi}
\newcommand\cmss{\ifmmode{\rm cm\thinspace s^{-2}}\else cm\thinspace s$^{-2}$\fi}
\newcommand\msun{\ifmmode{M_{\odot}}\else $M_{\odot}$\fi}
\newcommand\rsun{\ifmmode{R_{\odot}}\else $R_{\odot}$\fi}
\newcommand\mjup{\ifmmode{M_{\rm J}}\else $M_{\rm J}$\fi}
\newcommand\rjup{\ifmmode{R_{\rm J}}\else $R_{\rm J}$\fi}

\begin{document}

\bibliographystyle{apj}

\title{Mass and radius determinations for five transiting M-dwarf
stars.}

\author{Jose M. Fernandez\altaffilmark{1,2}, 
David W. Latham\altaffilmark{1}, 
Guillermo Torres\altaffilmark{1}, 
Mark E. Everett\altaffilmark{3}, 
Georgi Mandushev\altaffilmark{4}, 
David Charbonneau\altaffilmark{1}, 
Francis T. O'Donovan\altaffilmark{5},
Roi Alonso\altaffilmark{6},
Gilbert A. Esquerdo\altaffilmark{1}, 
Carl W. Hergenrother\altaffilmark{7}, 
Robert P. Stefanik\altaffilmark{1}
}

\altaffiltext{1}{Harvard-Smithsonian Center for Astrophysics, 60
Garden Street, Cambridge, MA 02138, USA; jfernand@cfa.harvard.edu.}

\altaffiltext{2}{Department of Astronomy, Pontificia Universidad
Cat\'{o}lica, Casilla 306, Santiago 22, Chile.}

\altaffiltext{3}{Planetary Science Institute, 1700 East Fort Lowell
Road, Suite 106, Tucson, AZ 85719, USA.}

\altaffiltext{4}{Lowell Observatory, 1400 W. Mars Hill Rd., Flagstaff,
AZ 86001, USA.}

\altaffiltext{5}{NASA Postdoctoral Program Fellow, Goddard Space
Flight Center, 8800 Greenbelt Rd Code 690.3, Greenbelt, MD 20771, USA.}

\altaffiltext{6}{Laboratoire d'Astrophysique de Marseille (UMR 6110),
Technople de Marseille-Etoile, F-13388 Marseille cedex 13, France.}

\altaffiltext{7}{Lunar and Planetary Laboratory, University of
Arizona, Tucson, AZ 85719, USA.}

\begin{abstract}
We have derived masses and radii for both components in five
short-period single-lined eclipsing binary stars discovered by the
TrES wide-angle photometric survey for transiting planets. All these
systems consist of a visible F-star primary and an unseen M-star
secondary ($M_{\rm A} \geq 0.8 \msun$, $M_{\rm B} \le 0.45
\msun$). The spectroscopic orbital solution combined with a high
precision transit light curve for each system gives sufficient
information to calculate the density of the primary star and the
surface gravity of the secondary. The masses of the primary stars were
obtained using stellar evolution models, which requires accurate
determinations of metallicities and effective temperatures. In our
case, the uncertainty in the metallicity of the primary stars is the
most important limiting factor in order to obtain accurate results for
the masses and radii of the unseen M-dwarf secondaries. The solutions
were compared with results obtained by calculating the radius of the
primary stars under the assumption of rotational synchronization with
the orbital period and alignment between their spin axis and the axis
of the orbit, using the observed broadening of the spectral lines as
an indicator of stellar rotation. Four systems show an acceptable
match between the two sets of results when their metallicity is
allowed to vary around solar values ($-0.5 \le \left[{\rm Fe/H}\right]
\le +0.5$), but one system shows a clear mismatch between the two
solutions, which may indicate the absence of synchronization or a
misalignment between the rotational and orbital axis.  When compared
to low-mass stellar evolution models, the derived masses and radii of
the unseen M dwarfs are inconsistent (three only marginally) with the
predicted values, with all of the radii being larger than expected for
their masses. These results confirm the discrepancy shown in previous
work between the predicted and observed radii on low-mass binary
stars. This work also shows that reliance on the assumption of
synchronization to derive the mass and radius of stars in eclipsing
single--lined F+M binaries is a useful tool, but may not always be
warranted and should be carefully tested against stellar evolution
models.
\end{abstract}


\section{Introduction}
\label{intro}
Dynamical determination of fundamental properties of stars has played
a key role in our understanding of stellar physics. Through careful
observations of eclipsing double-lined spectroscopic binaries it is
possible to obtain results with a precision as high as 1\% on their
masses and radii, or even better in some cases. These quantities
provide strong constraints on stellar evolution models. Low-mass stars
($M \le 0.5 \msun$) are particularly difficult to study, mainly
because of the low probability of finding eclipsing systems and also
due to their very low intrinsic brightness. Precise constraints on
their masses and radii are essential to model changes in their
equation of state, which goes from being close to an ideal gas for the
more massive stars, to a partially degenerate electron gas for M
dwarfs \citep{1998A&A...337..403B, 2000ApJ...542..464C}.  Thus, stars
near the bottom of the main sequence pose a real challenge to stellar
astronomers.

To date there are only eight M dwarfs (in four binary systems) with
published masses between 0.2 and 0.6 $\msun$ that have their masses
and radii measured to better than 3 percent: CM Draconis \citep{
2008arXiv0810.1541M}, YY Geminorum \citep{2002ApJ...567.1140T}, CU
Cancri \citep{2003A&A...398..239R} and GU Bootis
\citep{2005ApJ...631.1120L}. Remarkably, all these stars are
consistently larger than predicted by low-mass stellar evolution
models, and recent results point towards induced stellar activity as
the cause of the larger radii observed in low-mass double-lined
eclipsing binaries \citep{2007ApJ...660..732L,2007A&A...472L..17C}.

Because the numbers are so small, every additional well studied
low-mass star is important in our understanding of their fundamental
properties. In recent years, the growing number of short-period
single-lined eclipsing binaries (SEBs) with F-star primaries and
M-dwarf secondaries (hereafter F+M binaries) identified by photometric
surveys for transiting planets promises to provide a way to fill the
low-mass domain gap in the mass-radius diagram
\citep{2005A&A...431.1105B, 2005A&A...438.1123P, 2005A&A...433L..21P,
2006A&A...447.1035P, 2007ApJ...663..573B}. Their large amplitude
orbital radial velocities (several \kms) and transit depths between
1\% and 4\% make their follow up a reasonable task for 1--2-m class
telescopes on stars brighter than 12$^{\rm th}$ mag. However, unlike
double-lined eclipsing binaries, these systems do not directly provide
masses and radii because the radial velocity curve can be only
measured for the primary star (since only spectral lines from that
star are visible). Only the mean stellar density for the primary and
the surface gravity for the secondary star can be obtained directly
from the observables.

One way to obtain a full solution for the masses and radii of a
single--lined system is to use stellar evolution models to estimate
the mass of the primary star based on its luminosity, effective
temperature and metallicity. The same approach has been used to
calculate the mass and radius of transiting extrasolar planets, which
deliver uncertainties typically between 5\% and 10\% for the masses,
depending on how well constrained the atmospheric properties of the
primary stars are.

However, a different method can be used to solve for the masses and
radii of SEBs, which is virtually independent of stellar evolution
models. In F+M binaries, the gravitational interaction between the
primary and secondary star is at least two orders of magnitude
stronger than in planetary systems. For binaries with short orbital
periods ($\lesssim$ 6 d) and measured eccentricities close to zero,
one may in principle assume that two processes have already taken
place due to tidal forces: synchronization between the orbital motion
and stellar rotation, and alignment between the orbital and rotational
axis \citep{1977A&A....57..383Z, 1981A&A....99..126H}. The timescales
for synchronization and axis alignment can be 50 or 100 times shorter
than the time it takes to circularize the orbits, which can range from
less than 100 Myr for very short period binaries (1--2 d) to more than
1 Gyr for binaries with longer periods (5--6 d). If synchronization
and alignment have taken place, the radius of the primary star can be
obtained by combining the rotational velocity derived from the
observed broadening of the spectral lines with the orbital period and
the inclination of the orbit, setting the scale of the system. This
approach depends on the predictions from stellar models only in minor
ways: limb darkening coefficients are needed for the detailed analysis
of the eclipse light curve, and the rotational broadening that is
derived from the observed spectra can depend weakly on the metallicity
that is adopted. A precision (but not necessarily accuracy) of 5\% or
better in the mass and radius is possible if radial and rotational
velocities are good to 2\% and transit photometry is good to 1\%, as
will be shown in this work.

Interestingly, it has been noted previously in the literature that the
assumption of synchronization does not always seem to hold when tested
in detail. Two short-period low-mass transiting M dwarfs discovered by
OGLE \citep{2005A&A...433L..21P, 2006A&A...447.1035P} delivered
unrealistic properties for the primary stars if synchronization or
pseudo synchronization was assumed. Their mass and radius had to be
obtained from an estimation of the primary star properties using
stellar isochrones, with final errors near 10\% for the masses and 7\%
for the radii.

The goal of this paper is to contribute to our understanding of the
fundamental properties of low mass stars. To accomplish this, we have
determined the mass and radius for the components of five eclipsing
single-lined F+M binaries identified by the TrES wide-angle transiting
planet survey \citep[TrES,][]{2004ApJ...613L.153A}.  The next section
describes the discovery and follow up observations of our targets. In
section \ref{data_model} we describe the models that were fitted to
the data. In section \ref{results_discussion} we present our results
and discuss some implications of this study, in particular for the
dynamical evolution of short period binary stars, and the usefulness
of these systems to obtain a precise characterization of low-mass
stars.

\section{Observations and Data Reduction}
\label{observations}

\subsection{Detection through TrES Photometry}

Transit events for the systems studied in this paper where detected
early on during regular operations of the Trans-Atlantic Exoplanet
Survey network (TrES), an arrangement of three 10-cm telescopes
distributed in longitude. The three network nodes are: the STARE
telescope (Observatorio del Teide of the Instituto de Astrof\'isica de
Canarias, Spain), the Sleuth Telescope (Palomar Observatory,
California, USA), and the Planet Search Survey Telescope (Lowell
Observatory, Arizona, USA). More than 30 fields of 5.7$^{\circ}$
$\times$ 5.7$^{\circ}$ were monitored between the years 2003 and
2008. The data from each telescope were processed separately, as
described by \citet{2004PASP..116.1072D}. The binned light curves were
analyzed using the box-fitting transit search algorithm of
\citet{2002A&A...391..369K} to find periodic signals consistent with
the passage of a Jupiter-sized object across the disk of a solar-like
star. Hundreds of objects were flagged as planetary candidates because
of their shallow transit depth and the lack of a secondary eclipse or
significant out-of-transit photometric variations. In general, these
preliminary candidates were expected to fall within three categories:
transiting planets, photometric false detections, and astrophysical
false positives. Of all these candidates we selected five for the
present work, based on their spectroscopic properties and the
high-precision follow-up observations we were able to gather, as
described below. Coordinates, visual magnitudes, and near-infrared
colors for these five objects are given in Table \ref{coord_mag}.

\subsection{Follow-up Spectroscopy} 

\begin{figure}
\plotone{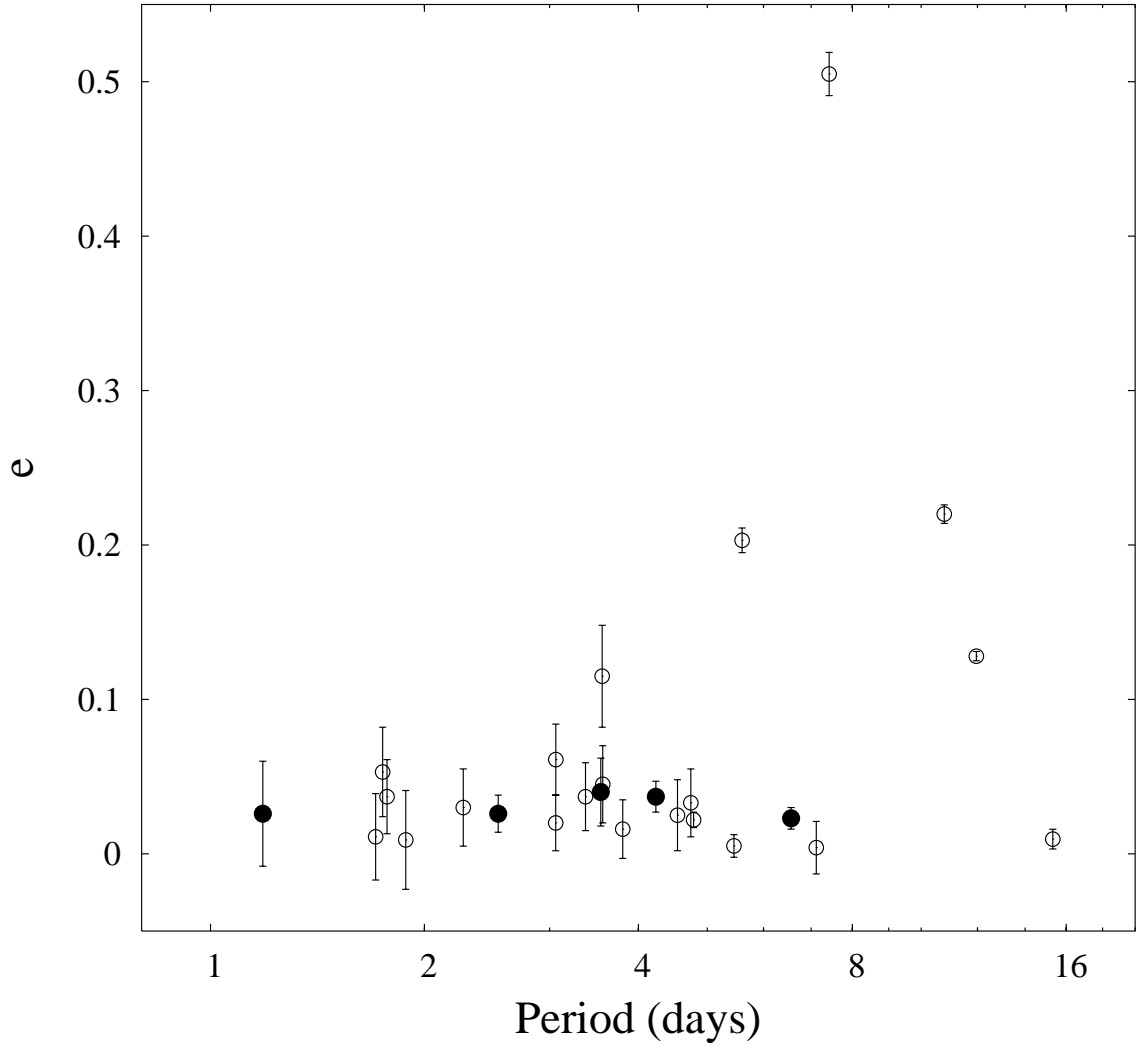}
\caption{ 
Eccentricity vs. orbital period diagram for 26 single--lined eclipsing
binary stars discovered by TrES. The five systems presented in this
work are shown as filled circles.
\label{fig1}}
\end{figure}

\begin{figure}
\plotone{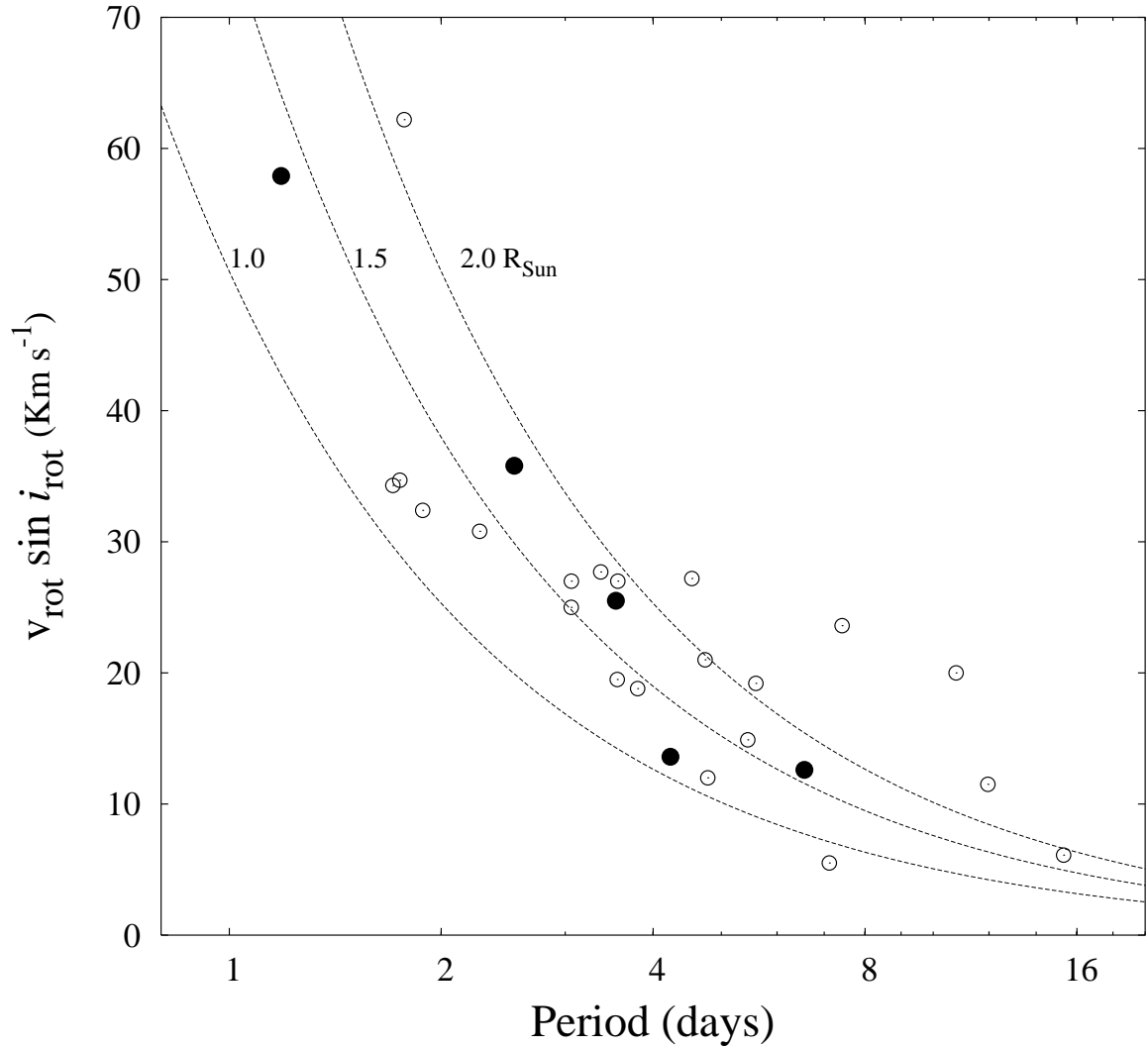}
\caption{
Rotational velocity vs. orbital period for 26 SEB's discovered by
TrES. Curves of constant radius in solar units are shown, computed
under the assumption of orbit--rotation synchronization. The five
systems studied in this work are shown as filled circles.
\label{fig2}}
\end{figure}

\begin{figure}
\plotone{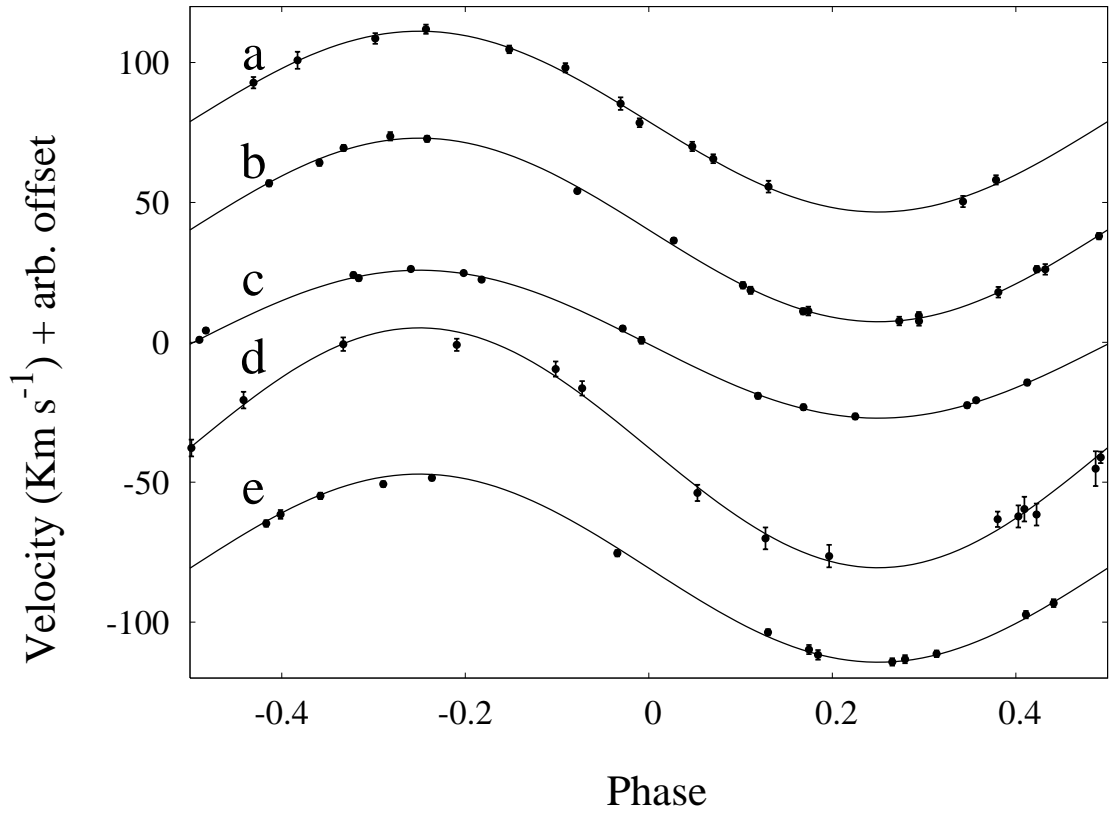}
\caption{ 
Period-phased radial velocities for T-Aur0-13378 (a),
T-Boo0-00080 (b) T-Lyr1-01662 (c), T-Lyr0-08070 (d) and T-Cyg1-01385
(e). Continuous lines show the best orbital  fit for each data set.
\label{fig3}}
\end{figure}

A common strategy for following up transiting-planet candidates
identified by wide-field photometric surveys is to start with an
initial spectroscopic reconnaissance, to see if there is evidence for
a stellar companion that might be responsible for the observed light
curve. For this we used the CfA Digital Speedometers
\citep{1992ASPC...32..110L} on the 1.5-m Wyeth Reflector at the Oak
Ridge Observatory in the town of Harvard, Massachusetts, USA and on
the 1.5-m Tillinghast Reflector at the Fred L.\ Whipple Observatory on
Mount Hopkins, Arizona, USA. We obtained single-order echelle spectra
in a wavelength window of 45 \AA\ centered at 5187 \AA, with a
resolution of 8.5 \kms, and a typical signal-to-noise ratio per
resolution element of 15 to 20.  For slowly rotating solar-type stars
these spectra deliver radial velocities accurate to about 0.5 \kms,
which is sufficient to detect orbital motion due to companions with
masses down to about 5 or 10 \mjup\ for orbital periods of a few
days. A detailed description of the spectroscopic data reduction can
be found elsewhere \citep{2004ApJ...613L.153A, 2007ApJ...663..573B,
2008arXiv0812.1161L}. From the first set of spectra it is clear when
candidates are not planets. If large variations in the radial
velocities are observed, the explanation for the photometric signal is
usually an eclipsing binary star (both signals should have a
consistent periodicity and phasing). A total of 26 orbital solutions
were derived for SEBs from the spectroscopic observations of planetary
candidates. Orbital periods for these systems range from 1.2 days to
15.3 days, and eccentricities range from 0.0 to 0.5 (see Figure
\ref{fig1}).  The rotational broadening of the spectral lines, to be
described in Section \ref{vsini_teff_fe}, also varies significantly
reaching $V_{\rm rot}\sin i_{\rm rot}$ values of up to 60 \kms \ or
more (see Figure \ref{fig2}). The phased radial velocities of the five
SEBs studied in this work are shown in Figure \ref{fig3}, and the
individual radial velocity measurements are presented in Table
\ref{rv_T-Aur0-13378} to Table \ref{rv_T-Cyg1-01385}. Table
\ref{rv_results} gives a comprehensive summary of the orbital
solutions obtained from the spectroscopy.

\subsection{Follow-up KeplerCam Photometry}

\begin{figure}
\plotone{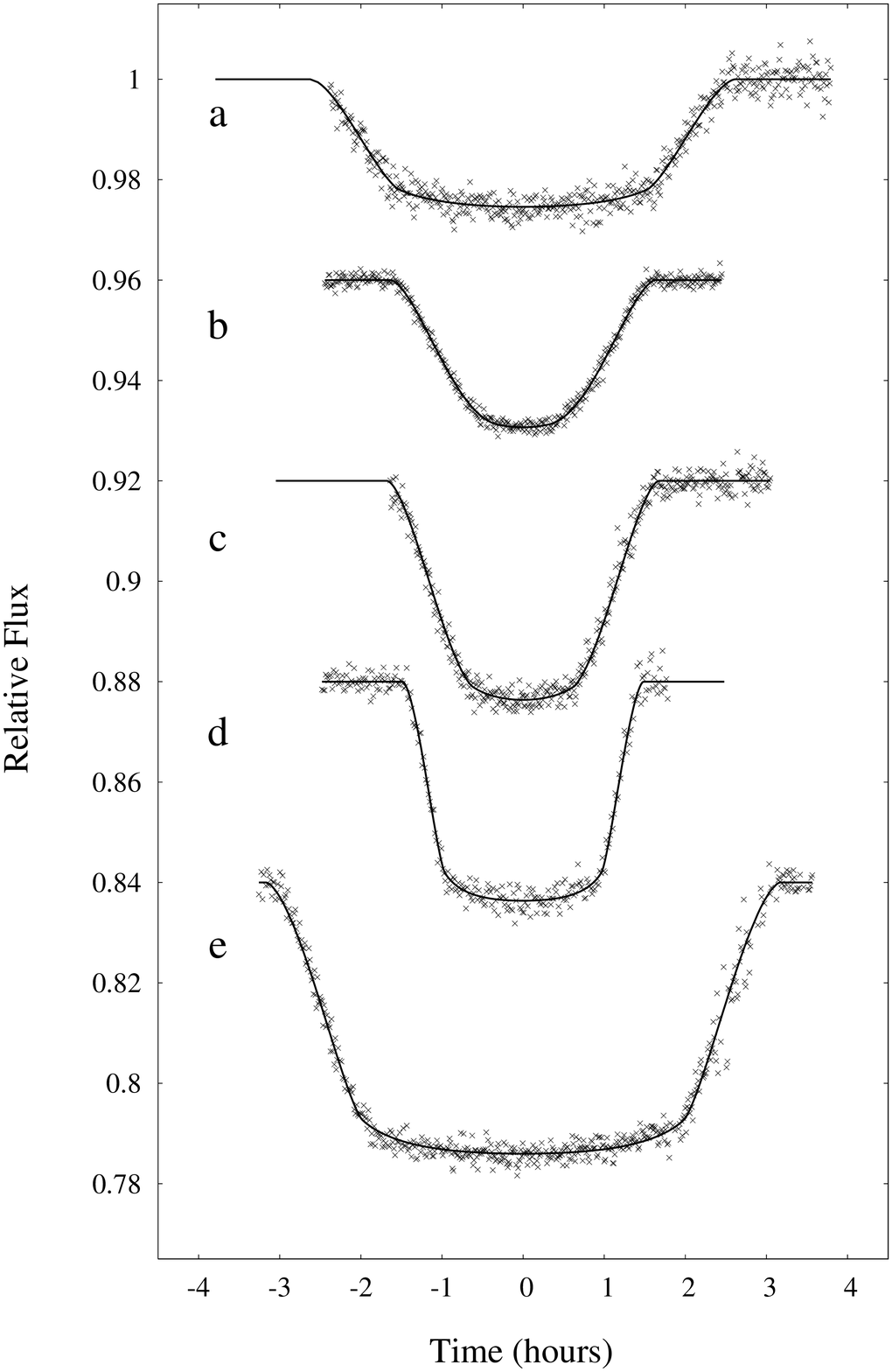}
\caption{
KeplerCam primary transit light curves for T-Aur0-13378 (a),
T-Boo0-00080 (b) and T-Lyr1-01662 (c) in the SDSS $i$ band, and
T-Lyr0-08070 (d) and T-Cyg1-01385 (e) in the $z$ band. Continuous
lines show the best fit for each data set.
\label{fig4}}
\end{figure}

In single-lined eclipsing systems is not possible to perform direct
measurements of the radius of the secondary component. The only
closely related observable quantity is the radius ratio between the
objects, obtained through careful analysis of the photometric transit
of the system. The main difficulty in obtaining a complete, high
signal-to-noise transit light curve is the necessity of continuous
excellent weather during several hours of observation. Because of this
limitation, we scheduled times-series photometric observations only
for those systems in which the assumption of orbit-rotation
synchronization seemed to be secure (short orbital periods, and
eccentricities close to zero), and with values of $V_{\rm rot}\sin
i_{\rm rot}$ larger than about 10 \kms, to avoid the accumulation of
larger errors in the determination of the radius of the primary star
as the spectroscopic resolution was approached. We used the predicted
eclipse times from our spectroscopic orbits to schedule observations
of the systems that passed this test, and successfully observed full
transits of five SEBs. To provide a high-quality light curve for the
analysis of the primary eclipse of each system, we used KeplerCam on
the 1.2-m telescope at the Fred L.\ Whipple Observatory on Mount
Hopkins, Arizona. KeplerCam utilizes a monolithic 4K$\times$4K
Fairchild 486 CCD that gives a $23 \arcmin \times 23 \arcmin$ field
and a pixel size of $0.68 \arcsec$ when the binning is $2 \times
2$. To minimize limb darkening effects on the shape of the transit
light curves, observations were made using long wavelength filters
(Sloan $i$ and $z$ bands). Relative aperture photometry was performed
to obtain the light curves. We iteratively selected comparison stars
by removing any that showed unusual noise or variability. The typical
RMS residual for the five light curves varied between 0.0010 and
0.0018 in relative flux units. Table \ref{kepcam_phot} gives a summary
of the photometric observations, with information about dates, pass
bands, exposure times, cadence, air-mass, FWHM and RMS. Figure
\ref{fig4} shows the observed light curves in scaled relative flux,
and samples of the observations are listed in Table
\ref{kepcam_lc}. We intend for this table to appear in its entirety in
the electronic version of the journal.

\subsection{Rotational Velocities, Effective Temperatures, and Metallicities}
\label{vsini_teff_fe}

\begin{figure}
\plotone{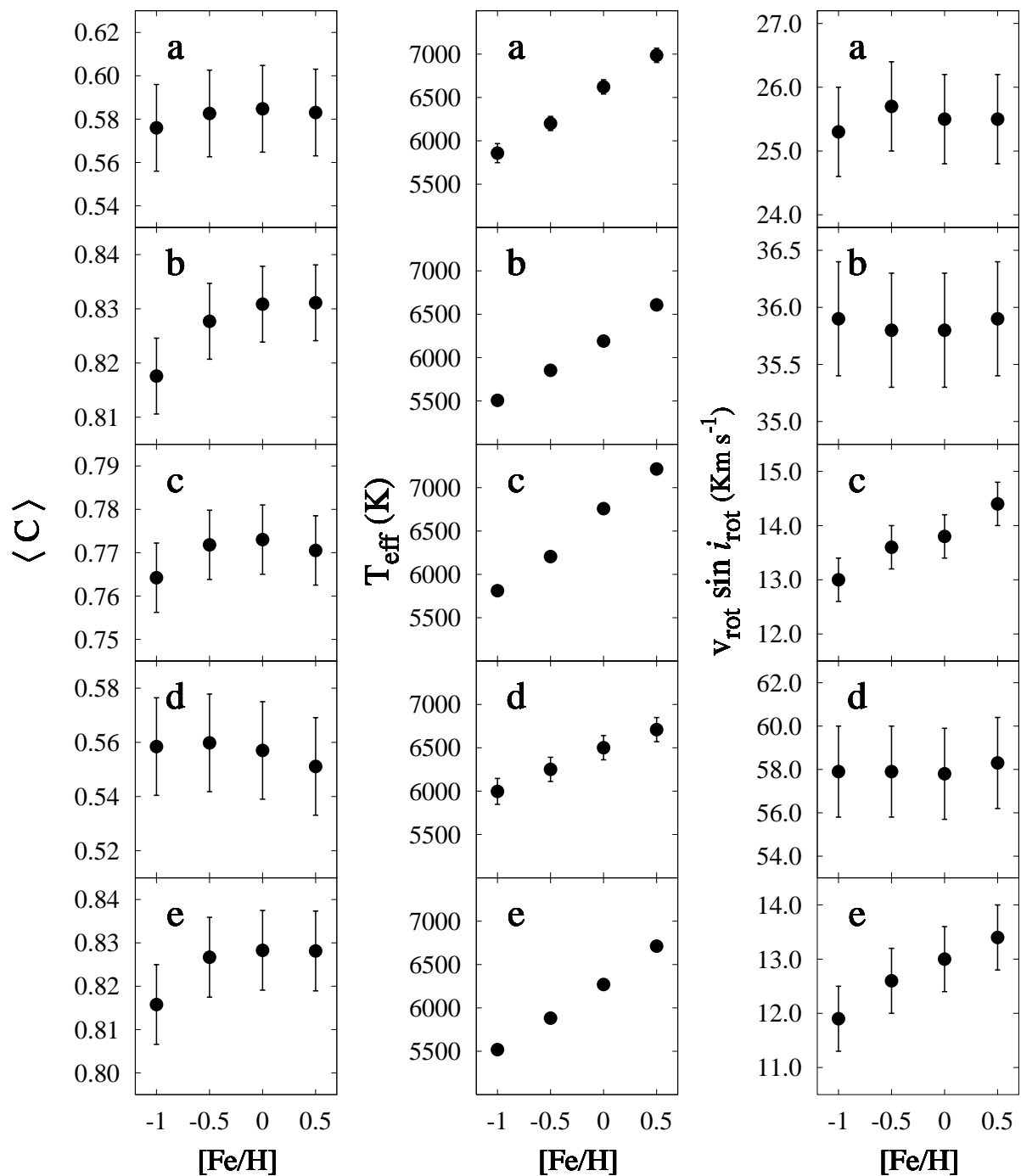}
\caption{ 
Dependence of the cross-correlation index, temperature and projected
rotational velocity on the adopted metallicity for T-Aur0-13378 (a),
T-Boo0-00080 (b) T-Lyr1-01662 (c), T-Lyr0-08070 (d) and T-Cyg1-01385
(e).
\label{fig5}}
\end{figure}

The key to obtaining masses and radii under the assumption of
synchronization is the determination of $V_{\rm rot} \sin i_{\rm rot}$
($i_{\rm rot}$ being the inclination angle between the rotational axis
of the primary star and the line of sight), because when combined with
the orbital period and orbital inclination, it sets the size of the
primary star (assuming as well that $i_{\rm rot} = i_{\rm orb}$, the
latter being the angle between the orbital axis and the line of
sight). The projected rotational velocity, along with the effective
temperature and surface gravity of each star, were extracted from the
same echelle spectra used to derive radial velocities. We
cross-correlated our spectra against a library of synthetic spectra
created by J. Morse using Kurucz model atmospheres \citep[see,
e.g.,][]{2002AJ....124.1144L} to estimate these properties for the
primary stars for a given metallicity, which we fixed at four
different values: $\left[{\rm Fe/H}\right] \ = \ -1.0$, $-0.5$, $0.0$
and $+0.5$. For each spectrum we looked for the metallicity-indexed
model with the highest correlation, which gave a corresponding
rotational velocity, temperature and surface gravity. The adopted
values for correlation index $C$, projected rotational velocity
$V_{\rm rot}\sin i_{\rm rot}$ and effective temperature $T_{\rm eff}$
were obtained averaging over all the individual observations, after
performing a three-sigma rejection. The internal error is the observed
standard deviation divided by the square root of the number of
measurements. In most cases, the average correlation index $\langle
C\rangle$ is highest when solar metallicity is considered, but the
typical error is generally larger than the differences. Thus this
method cannot be used to obtain precise and accurate metallicities
(see left panels of Figure \ref{fig5}). Because of the narrow
wavelength coverage of our spectra, there is a clear dependence
between the stellar properties (effective temperature in particular)
and the adopted metallicity. To decrease the level of degeneracy, we
iteratively used the surface gravity of the primary star obtained at
the end of the procedure to be explained in Section \ref{iso_sol} as
an additional constraint for the cross correlation routine. Figure
\ref{fig5} shows the relation between metallicity, correlation index,
effective temperature and rotational velocity for the five systems
under study. Table \ref{atm_prop} gives the values and uncertainties
for the primary star atmospheric parameters for each adopted
metallicity.

\section{Data Modeling}
\label{data_model}

\subsection{Light Curve Analysis}

Modeling of the light curves was carried out employing the formalism
of \citet{2002ApJ...580L.171M}, using a quadratic limb-darkening law,

\begin{equation}
\label{quad_limb_dark}
I_{\mu} = 1 - u_{1}(1-\mu) - u_{\rm 2}(1-\mu)^{2}
\end{equation}
where $I_{\mu}$ is the observed intensity relative to the center of
the stellar disk and $\mu$ is the cosine of the angle between the line
of sight and the normal to the stellar surface. The limb darkening
coefficients $u_{1}$ and $u_{2}$ were taken from the tables of
\citet{2004A&A...428.1001C} adopting the metallicity-dependent values
of $T_{\rm eff}$ and $\log g$. The periods were held fixed and the
orbits were assumed to be circular since the eccentricities obtained
from the orbital solutions are close to zero for all systems. The
fitted parameters were the radius ratio $R_{\rm B}/R_{\rm A}$, the
reduced semi-major axis $a/R_{\rm A}$, and the impact parameter $b$,
defined as $b=a/R_{\rm A} \cos i_{\rm orb}$ where $ i_{\rm orb}$ is
the inclination of the orbital plane to the line of sight. Here
$R_{\rm A}$ and $R_{\rm B}$ correspond, respectively, to the radius of
the primary and secondary star in each binary. The same notation is
used for all other parameters in the text and equations that contain
the A and B sub-indexes.

The best fit between the model and the data was found minimizing
$\chi^{2}_{lc}$:

\begin{equation}
\label{chisq_lc}
\chi^{2}_{lc} = \sum_{i=1}^{N_{f}} \left[ \frac{f_{i}^{obs} - f_{i}^{mod}}{\sigma_{i}}\right]^{2} \ ,
\end{equation}
where $f_{i}^{obs}$ and $f_{i}^{mod}$ are the observed and modeled
relative fluxes observed at time $i$, and $\sigma_{i}$ is the error of
each data point. The best values and uncertainties for the fitted
parameters were obtained using a Markov Chain Monte Carlo simulation
(MCMC). As described by \citet{2005AJ....129.1706F} and
\citet{2006ApJ...652.1715H}, in this method a random process is used
to create a sequence of points in parameter space that approximates
the studied probability distribution. This sequence or chain is
generated by a jump function that adds a Gaussian random number to
each parameter. The jump is executed if the new point has a
$\chi^{2}_{lc}$ lower than the previous point. If $\chi^{2}_{lc}$ is
larger, the jump is made with a probability equal to
exp$\left(-\Delta\chi^{2}_{lc}\right)$. If the jump is not made, the
new point is a copy of the previous one. The relative sizes of the
perturbations were set using the uncertainties obtained by direct
inspection of $\chi^{2}_{lc}$ across the parameter space, as done in
\citet{2007ApJ...663..573B}. The sizes of the jumps are set by
requiring that $\sim$25\% of the jumps are accepted. Four independent
chains of 55000 points were created for each light-curve, starting
from a point 5-$\sigma$ away from the optimal values obtained by
direct inspection, and discarding the first 20\% of the points to
minimize initial condition effects. The four chains were combined to
create one long sequence of points. The best-fit value and
uncertainties for each parameter were obtained from the value interval
centered on the median that contains 68\% of the points (1--$\sigma$
errors). The results derived using limb-darkening coefficients
adopting solar metallicity are summarized in Table \ref{lc_fit}. The
differences between these results and those adopting different
metallicities are negligible compared to their errors.

\subsection{Density and Surface Gravity}

As shown previously by \citet{2003ApJ...585.1038S},
\citet{2004MNRAS.355..986S} and \citet{2007ApJ...664.1190S}, an
approximation to the mean stellar density of the primary star
($\rho_{\rm A}$) and the surface gravity of the secondary ($g_{\rm
B}$) can be derived directly from Newton's modified version of
Kepler's third law and the mass function of the binary. The familiar
expressions

\begin{equation}
\label{kepler_3rd}
a^{3} = \frac{G}{4 \pi^{2}}\left(M_{\rm A} + M_{\rm B}\right) P^2
\end{equation}

\begin{equation}
\label{mass_func}
M_{\rm B} = \left(\frac{P}{2 \pi G}\right)^{1/3} \ \frac{K_{\rm A}}{\sin i_{\rm orb}}\left(M_{\rm A} + M_{\rm B}\right)^{2/3}
\end{equation}
can be combined and re-written in the following form:

\begin{equation}
\label{density}
\rho_{\rm A} = \frac{3 \pi}{GP^2}\left(a/R_{\rm A}\right)^3 \ - \rho_{\rm B} \left(R_{\rm B}/R_{\rm A}\right)^3
\end{equation}

\begin{equation}
\label{surf_grav}
g_{\rm B} = \frac{2 \pi}{P} \ \frac{K_{\rm A}}{\sin i_{\rm orb}} \ \left(\frac{a/R_{\rm A}}{R_{\rm B}/R_{\rm A}}\right)^2
\end{equation}
where $\rho_{\rm B}$ is the density of the secondary star. The
expression for the mass function in eq. \ref{mass_func} assumes zero
eccentricity, which is a good approximation in our case. This
assumption is also supported by studies such as those of
\citet{1971AJ.....76..544L} and \citet{2005A&A...439..663L}, which
show that for single-lined binaries, small observed eccentricities ($e
< 0.1$) are generally not statistically significant.

Using the parameters obtained earlier from modeling the transit light
curves ($R_{\rm B}/R_{\rm A}$, $a/R_{\rm A}$, $b$), along with the
measured orbital parameters ($P$, $K_{\rm A}$), we may restrict the
location of the stars in each system on the mass-radius diagram to
unique curves of constant stellar density for the primary, and
constant surface gravity for the secondary, which are described
completely by the observables \citep[see,
e.g.][]{2007ApJ...663..573B}:

\begin{eqnarray}
\label{density_curve}
M_{\rm A} & = & \frac{4 \pi^{2}}{G P^2} \ \left(a/R_{\rm A}\right)^3 \  \left(1 - \frac{P \ K_{\rm A} \ }{2 \pi {\left(1-\left(b^2/\left(a/R_{\rm A}\right)^2\right)\right)}^{1/2}\ \left(a/R_{\rm A}\right) \ R_{\rm A}}\right) \ R_{\rm A}^3 \\
\label{surf_grav_curve}
M_{\rm B} & = & \frac{2 \pi}{G P} \left(\frac{a/R_{\rm A}}{R_{\rm B}/R_{\rm A}}\right)^2 \frac{K_{\rm A}}{\left(1-\left(b^2/\left(a/R_{\rm A}\right)^2\right)\right)^{1/2}}\ R_{\rm B}^2
\end{eqnarray}
where $\sin i_{\rm orb}$ has been re-written in terms of the
observables $a/R_{\rm A}$ and $b$. These expressions are essentially
model-independent and are very useful for the following reason: if an
independent measurement of the mass or the radius of the primary star
(or secondary star) is made, it is possible to calculate a full
solution for the system (making use of the radius ratio $R_{\rm
B}/R_{\rm A}$).

\subsection{Mass and radius determinations}

As mentioned in Section \ref{vsini_teff_fe}, there is a clear
dependency between the adopted metallicity of the primary star and its
atmospheric properties derived from our spectra, particularly the
effective temperature. A metallicity close to solar gave the best
match between the observed spectra and the models in most cases, but
it is not possible to independently obtain an accurate abundance for
these stars from our data alone. To address this problem, in the
following we model the data for the five systems assuming four
different metallicities: $\left[{\rm Fe/H}\right] \ = \ -1.0, -0.5,
0.0, +0.5$.

\subsubsection{System solution from stellar isochrones}
\label{iso_sol}

In order to obtain the mass and radius for the unseen M dwarf, we
estimated the mass of the primary star using the Yonsei-Yale stellar
evolution models \citep{2001ApJS..136..417Y, 2004ApJS..155..667D},
following the procedure of \citet{2008ApJ...677.1324T}. For this
purpose we relied on the adopted metallicity and the temperature
obtained from the spectra.  Parallaxes have not been measured for
these targets, so instead of the luminosities of the stars, we used
the parameter $a/R_{\rm A}$ derived from the modeling of the light
curves, which is closely related to the mean stellar density.

The observed quantity $a/R_{\rm A}$ can be compared directly to the
models re-writing eq. \ref{kepler_3rd}:

\begin{equation}
a/R_{\rm A} = \left(\frac{G}{4\pi^{2}}\right)^{1/3} \ \frac{P^{2/3}}{R_{\rm A}} \ \left(M_{\rm A} + M_{\rm B}\right)^{1/3}
\end{equation}

The mass of the secondary star is not know $a \ priori$, but an
approximate value sufficient for the present purpose can be estimated
using eq. \ref{surf_grav_curve} and the isochrones from
\citet{1998A&A...337..403B} given that the age dependency is weak for
low-mass stars. Once the secondary star mass has been derived, the
process can be repeated until convergence.

The Yonsei-Yale isochrones were interpolated to a fine grid in
metallicity and age and compared point by point with the measured
values of $T_{\rm eff}$ and $a/R_{\rm A}$.  As noted previously, the
range of metallicities explored was $\left[{\rm Fe/H} \right]=-1.0,
-0.5, 0.0, +0.5$. We adopted an arbitrary error of $\sigma_{\left[{\rm
Fe/H}\right]}=\pm0.2$, meant to illustrate the behavior of the results
over a wide range of metallicities in a compact way. The internal
error for $T_{\rm eff}$ was increased to account for the dependency
with $\left[{\rm Fe/H}\right]$ in the mentioned range, from the
nominal precision at a given metallicity of about 50 K to a more
conservative 200 K.

Each point on the isochrones that was consistent with $\left[ {\rm
Fe/H}\right]$, $T_{\rm eff}$ and $a/R_{\rm A}$ within their errors was
recorded and a likelihood given by $L = \exp\left( -\chi^2_{iso} /
2\right)$ was calculated for it, where

\begin{equation}
\label{chisq_iso}
\chi^2_{iso} = \left(\frac{\Delta \left[{\rm Fe/H}\right]}{\sigma_{\left[{\rm Fe/H}\right]}}\right)^{2} \ + \ \left(\frac{\Delta T_{\rm eff}}{\sigma_{T_{\rm eff}}} \right)^{2} \ + \ \left(\frac{\Delta \left( a/R_{\rm A}\right)}{\sigma_{a/R_{\rm A}}} \right)^{2} \ .
\end{equation}

The $\Delta$ symbols represent the difference between the observed and
model values for each quantity. The possible values for $L$ range
between $1$ (an exact match between observations and models) and
$0.22$ (the worst acceptable match). The best fit value for each
stellar parameter was obtained by adding all matches, weighted by
their corresponding normalized likelihood $L$.  The adopted errors for
the fitted parameters (mass and age) come from their range among the
accepted points on the isochrones. In the case of solar metallicity,
primary masses range from 1.3 \msun\ to 1.6 \msun, with uncertainties
ranging from 8\% to 13\%. Ages range from 0.7 Gyr to 4.0 Gyr, with
large uncertainties, from 25\% to 100\% in some cases. For lower
metallicities, more massive and older stars are obtained. The opposite
effect is observed for higher abundances.

With $M_{\rm A}$ known, we then obtained $M_{\rm B}$ and $a$ by
iteration of equations \ref{kepler_3rd} and \ref{mass_func}. The value
of $a$ in combination with $a/R_{\rm A}$ and $R_{\rm B}/R_{\rm A}$
allowed us to obtain consistent values for $R_{\rm A}$ and $R_{\rm
B}$. At this point, we used the newly derived surface gravity of the
primary star $\log g_{\rm A}$ as an additional constraint in the
determination of $\langle C\rangle$, $T_{\rm eff}$ and $V_{\rm rot}
\sin i_{\rm rot}$ as explained in Section \ref{vsini_teff_fe}. We
iterated this procedure until convergence.

To estimate the errors for $R_{\rm A}$, $R_{\rm B}$, and $M_{\rm B}$,
we used the MCMC chains generated in the course of modeling the
transiting light curves. For each element of the chain a solution was
calculated using the corresponding values of $R_{\rm B}/R_{\rm A}$,
$a/R_{\rm A}$ and $b$ together with random values for $P$ and $K_{\rm
A}$, normally distributed around the observed values with $\sigma$
equal to the measured uncertainties. In this way we obtained a
probability distribution for the masses and radii, from which we
extracted the median and 68\% confidence limits (1--$\sigma$) and
adopted them as best values and errors, respectively.

For solar metallicity, secondary masses range from 0.28 \msun\ to 0.43
\msun, and secondary radii range from 0.28 \rsun\ to 0.40 \rsun. The
uncertainties in $M_{\rm B}$ range from 5\% to 8\%, and those in
$R_{\rm B}$ from 3\% to 5\%. For sub-solar composition the masses and
radii are smaller, and the opposite effect is observed for higher
abundances.

\begin{figure}
\centering
\includegraphics[width=0.8\textwidth]{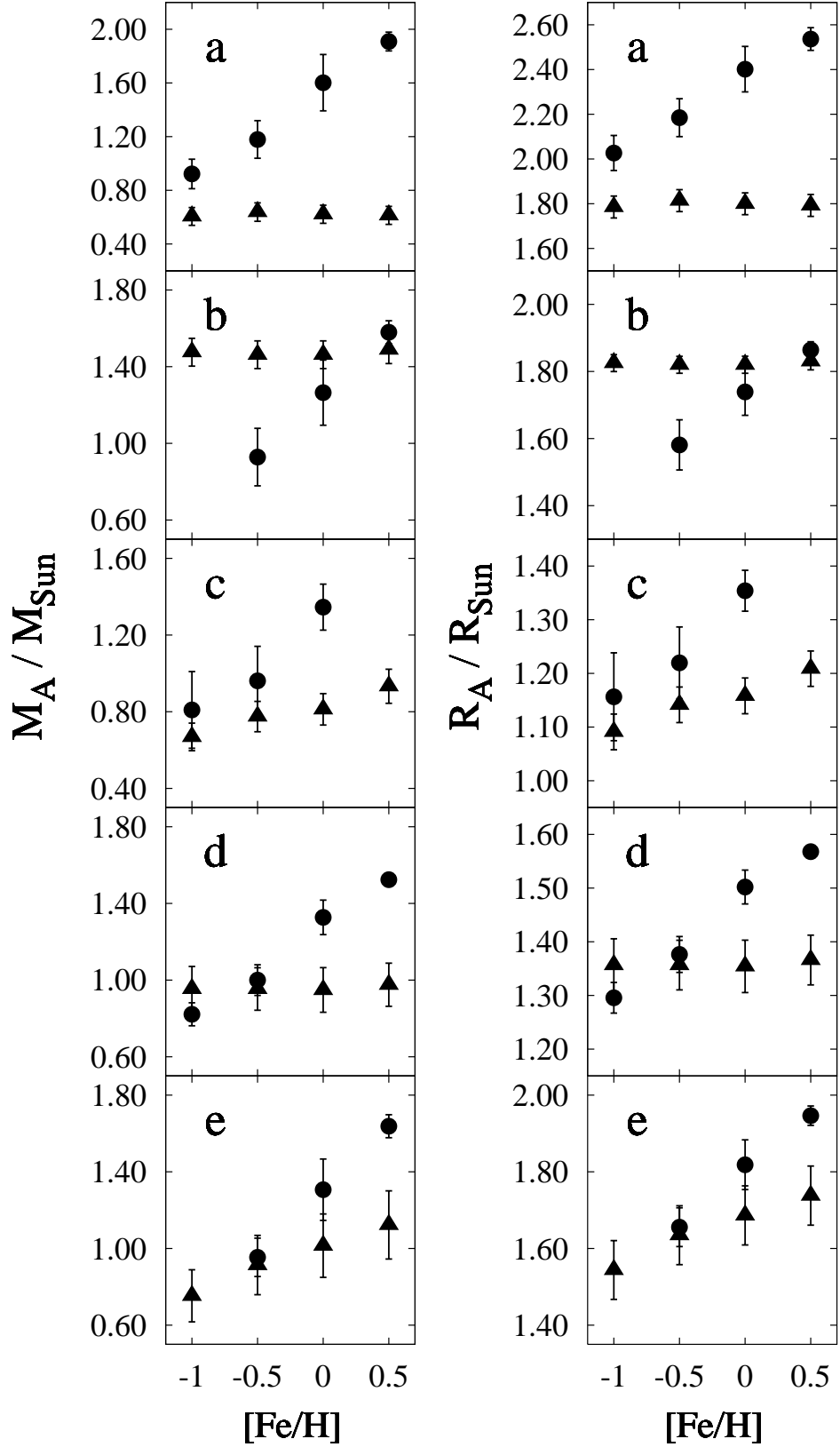}
\caption{ 
Dependence of the mass and radius of the primary star on the adopted
metallicity, for T-Aur0-13378 (a), T-Boo0-00080 (b) T-Lyr1-01662 (c),
T-Lyr0-08070 (d) and T-Cyg1-01385 (e). Results derived from stellar
isochrones are shown as filled circles, while the results derived
under the assumption of orbit-rotation synchronization are shown as
filled triangles.}
\label{fig6}
\end{figure}

The resulting masses and radii from the isochrone modeling of the
primary stars are shown in Figure \ref{fig6} with filled circles. For
some cases of very high or low metallicity, no consistent solution was
found based on stellar models, which explains some missing results in
the figures. In these cases, the value for $\log g$ from the nearest
consistent solution was used to better constrain the atmospheric
properties of the primary star (Section \ref{vsini_teff_fe}).

\subsubsection{System solution from orbit-rotation synchronization}
\label{rot_sol}

The uncertainty in the determination of metallicities for the primary
stars has significant consequences when estimating their masses using
stellar models. To overcome this serious limitation, we computed
masses and radii based on the assumption of synchronization and
co-alignment between the primary star's rotational axis and the axis
of the orbit.

From the rotational velocity, orbital period, and orbital inclination,
it is possible to calculate the radius of the primary star $R_{\rm
A}$ using the expression  

\begin{equation}
\label{RA_sync}
R_{\rm A} =  \frac{P}{2\pi \left(1-\left(b^2/\left(a/R_{\rm A}\right)\right)^2\right)^{1/2}}\ \left(V_{\rm rot} \sin i_{\rm rot}\right)
\end{equation}
which is the generalization of the geometric relation $P_{\rm
rot}=\frac{2\pi R}{V_{\rm rot}}$ for an impact parameter $b$ other
than zero, under the assumption of synchronization ($P=P_{\rm
orb}=P_{\rm rot}$) and alignment between the rotation and orbital axes
($i_{\rm orb}=i_{\rm rot}$). The radius ratio $\left(R_{\rm B}/R_{\rm
A}\right)$ is then used to infer the size of secondary star. Once both
radii are known, they can be incorporated into equations
\ref{density_curve} and \ref{surf_grav_curve} to obtain full
expressions for the masses.

For the error propagation we adopted uncertainties in $V_{\rm rot}
\sin i_{\rm rot}$ as described in Sect. \ref{vsini_teff_fe}. We
verified that for each of our five systems this error was larger than
the variations that come from the correlation between $V_{\rm rot}
\sin i_{\rm rot}$ and $\left[{\rm Fe/H}\right]$ over the range of
metallicities considered here. The final values and errors for the
masses and radii of both components were obtained using the MCMC
distributions from the light curve fits, incorporating the errors of
$a/R_{\rm A}$, $R_{\rm B}/R_{\rm A}$, $b$, $P$, $V_{\rm rot} \sin
i_{\rm rot}$ and $K_{\rm A}$ in a way analogous to the procedure used
for the parameters inferred by using stellar isochrones.

For the primary stars, masses range from 0.8 \msun\ to 1.5 \msun\
(with errors between 5\% and 15\%) and radii range from 1.1 \rsun\ to
1.8 \rsun\ (errors of 2 -- 5\%). Among the secondary stars, masses
range from 0.2\ \msun\ to 0.35 \msun\ (with errors between 3\% and
10\%) and radii range from 0.25 \rsun\ to 0.35 \rsun\ (errors of 2 --
5\%). The masses and radii for the primary stars obtained from the
assumption of synchronization are shown in Figures \ref{fig6} with
filled triangles.

\section{Results and Discussion}
\label{results_discussion}

We begin by comparing the results obtained for the primary stars using
stellar isochrones with those from synchronization assumptions. Figure
\ref{fig6} shows masses and radii vs. metallicity for every system. We
remind the reader that the error bars in the results derived from
stellar models include the contribution from a conservative error in
the adopted metallicity of each star.

For every system but one (T-Aur0-13378) there is a metallicity range
where both sets of results are consistent, but outside this range the
solutions diverge considerably. The results derived from estimating
the mass of the primary star using models show a strong dependence on
the adopted metallicity, not only because of its direct impact in the
isochrones, but also because the effective temperature is highly
correlated with it. The results derived from orbit-rotation
synchronization have a small dependence on the adopted metallicity due
to the weak correlation between the observed rotational velocity of
the primary star and the adopted abundance.

Tidal theory predicts these kinds of systems (short period, near
circular orbit) should be synchronized and their orbital and
rotational axes co-aligned. Following the formalism of
\citet{1981A&A....99..126H}, we calculated the ratio of the orbital
and rotational angular momentum $\alpha = \frac{q}{\left(1+q\right)}
\frac{1}{r_{g}^{2}}\left(a/R_{\rm A}\right)^{2}$, where $q = M_{\rm B}
/ M_{\rm A}$ and $r_{g}$ is the gyration radius of the primary star
($r_{g}^{2} = \frac{I}{M_{\rm A}R_{\rm A}^{2}}$, with $I$ being the
moment of inertia). For our binaries, $\alpha$ is always larger than
$70$, which means that the timescales for synchronization and
alignment of the orbital and rotational axes are expected to be much
shorter than the timescale for circularization. In order to have
comparable timescales for the three processes, $\alpha$ must be
between $5$ and $10$ \citep{1981A&A....99..126H}.  A similar relation
between timescales was obtained by Zahn (1977), and again
synchronization is predicted to occur much more rapidly than
circularization for these binary systems.

Even when evidence in favor of synchronization is strong, its reality
is not always guaranteed. The F+M binary OGLE-TR-123, studied by
\citet{2006A&A...447.1035P} has a very short period (1.8 d) and a
circular orbit, but the solution obtained if synchronization is
assumed does not match the properties of the primary star derived
spectroscopically. In this case, the solution derived from
synchronization implies a primary star which is not massive enough
($M_{\rm A}<0.9 \msun$) to explain its observed high effective
temperature ($T_{\rm eff} \sim 6700 K$).

The same inconsistency affects the system T-Aur0-13378, with a clear
disagreement between the solutions based on the assumption of
synchronization and those derived from stellar models. The main
problem with the solution based on synchronization is the
unrealistically low mass and radius inferred for the primary star,
given the relatively high temperatures obtained from spectroscopy. In
other words, the measured rotational velocity is too low, resulting in
a small radius. There are various possible explanations for this
behavior. The primary star in this system has the lowest density,
lowest surface gravity and the highest mass for a given metallicity
(when using models), if compared to the other four (see Table
\ref{lc_fit} and Table \ref{iso_results}). If solar or
lower-than-solar metallicity is adopted for the modeling, the
resulting primary is an evolved F-star with $M_{\rm A} < 1.6\msun$ and
$R_{\rm A} > 2.0\msun$. If this is the case, it could be that
conservation of angular momentum during the expansion of the star has
slowed down the surface rotation, and tidal forces have not been able
to keep up. Also, the presence of a radiative envelope in the primary
could be responsible for making tidal forces less efficient.  The
discrepancy between the two solutions could also be explained by a
misalignment between the spin axis of the primary and the axis of the
orbit, which is, however, not expected from the same theory that tells
us that synchronization should be taking place. By taking advantage of
the Rossiter-McLaughlin effect \citep{1924ApJ....60...15R,
1924ApJ....60...22M} it should be possible to measure the projection
of this angle on the plane of the sky \citep[see,
e.g.,][]{2006ApJ...653L..69W}, but one would still need the orthogonal
projection along the line of sight to solve the orientation of the
system completely. One way to independently determine the rotational
period of the primary star would be to obtain high-quality light
curves and measure the photometric variation outside of eclipse that
might be produced by rotation and star spots. This would serve as a
check on the assumption of tidal synchronization in the system.
Unfortunately, F stars are likely to be too hot for spot activity to
be detectable in their light curves by ground-based facilities
\citep[see, e.g.,][]{1994MmSAI..65...73H}. Dedicated spaced-based
surveys like CoRot \citep{2006cosp...36.3749B} and Kepler
\citep{2003ASPC..294..427B, 2005NewAR..49..478B} may be able to detect
this signal in similar systems.

Inconsistent solutions between tidal theory and stellar evolution
similar to the case of T-Aur0-13378 would be obtained as well for the
systems T-Lyr1-01662, T-Lyr0-08070, and T-Cyg1-01385 if accurate
spectroscopic determinations revealed the primary star metallicities
to be greater than solar. The system T-Boo0-00080 behaves differently
from the others, in the sense that high metallicities are necessary to
match both set of results. If the primary star in this system has a
metallicity lower than solar, the measured rotational velocity would
be too high for the system to be synchronized, and could be evidence of
an ongoing tidal process where the primary star is slowing down from a
faster initial rotation.

In view of the importance of knowing accurately the metallicity of the
primary stars, we have chosen to tabulate the full range of results
obtained from their modeling using stellar isochrones corresponding to
four different abundances (Table \ref{iso_results}). Additionally, we
have tabulated the results derived from the assumption of
orbit-rotation synchronization that show the best match with those
derived from stellar models (Table \ref{sync_results}). Strictly
speaking, the errors for the masses and radii obtained from modeling
the primary star with stellar isochrones are partially arbitrary
because of the adopted error on the metallicities. The errors for the
results based on synchronization are realistic and are dominated by
the uncertainty in the measured rotational velocity.

\begin{figure}
\plotone{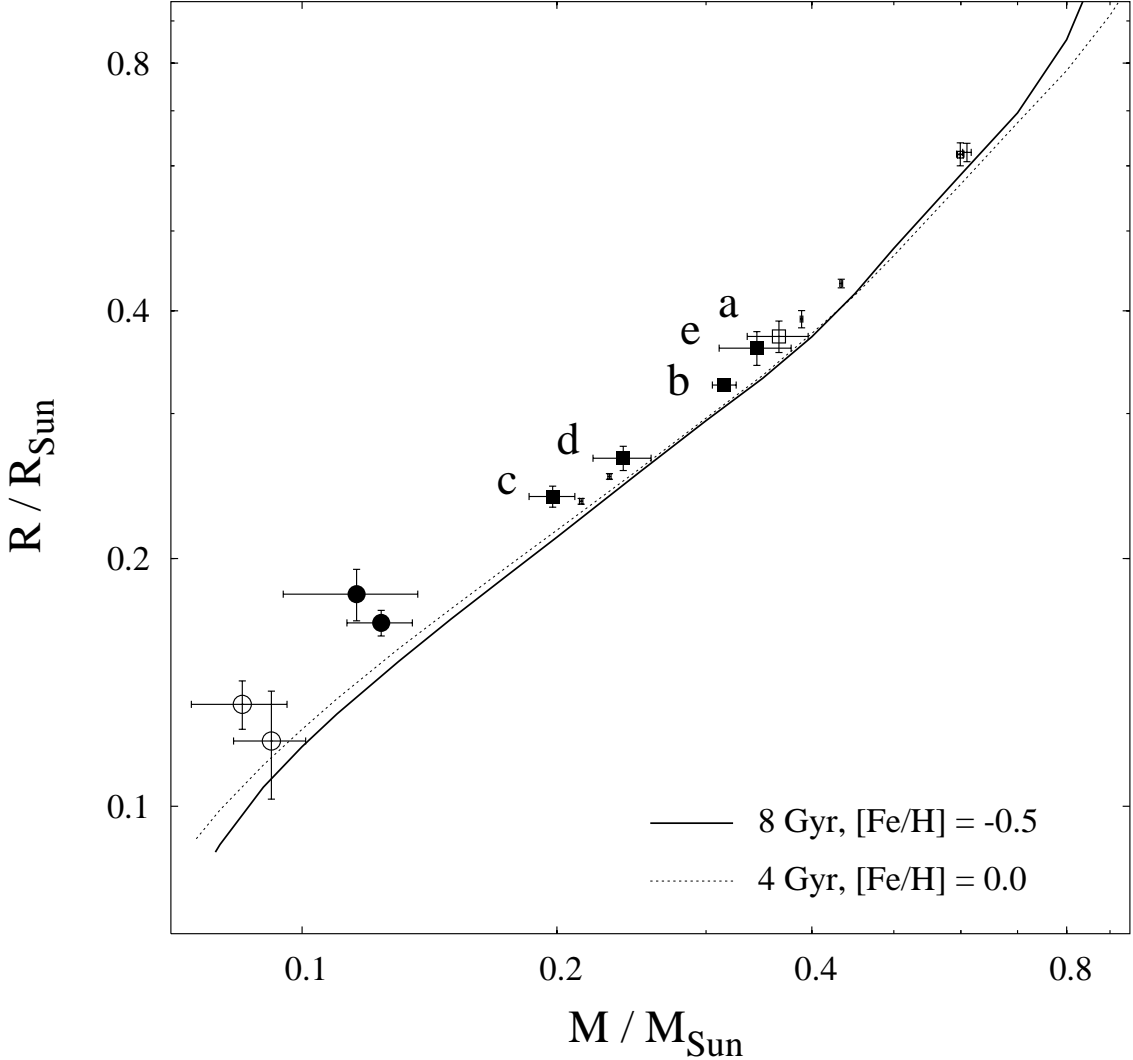}
\caption{
Mass-radius diagram for the M dwarfs T-Aur0-13378 (a), T-Boo0-00080-B
(b),T-Lyr1-01662-B (c), T-Lyr0-08070-B (d) and T-Cyg1-01385-B (e),
shown as squares. Four additional M dwarfs in SEBs studied by other
authors are shown as circles \citep{ 2005A&A...438.1123P,
2005A&A...433L..21P, 2006A&A...447.1035P, 2007ApJ...663..573B}. Filled
symbols correspond to results derived from the assumption of
orbit-rotation synchronization, and open symbols correspond to results
derived from stellar models. Eight M dwarfs in double-lined eclipsing
binaries (four systems total) are shown as dots \citep{
2008arXiv0810.1541M, 2002ApJ...567.1140T, 2003A&A...398..239R,
2005ApJ...631.1120L} together with low-mass stellar isochrones
\citep{1998A&A...337..403B}.
\label{fig7}}
\end{figure}

In Figure \ref{fig7} we display the mass and radius of the M-dwarf
secondaries of the systems T-Boo0-00080, T-Lyr1-01662, T-Lyr0-08070,
and T-Cyg1-01385 using the synchronization-based results from Table
\ref{sync_results}. The system T-Aur-013378 is shown using the
model-based results ($\left[{\rm Fe/H} \right]= 0.0 \pm 0.2$) because
its uncertain tidal configuration does not allow a reliable
determination of the mass and radius of the secondary star when
assuming orbit-rotation synchronization. We also show the results for
the M dwarfs HAT-TR-205-013 \citep{2007ApJ...663..573B}, OGLE-TR-106
\citep{2005A&A...438.1123P}, OGLE-TR-122 \citep{2005A&A...433L..21P},
and OGLE-TR-123 \citep{2006A&A...447.1035P}, which have the lowest
measured mass and radius available in the current literature for main
sequence stars. These four stars are also members of single-lined
spectroscopic binary systems. As mentioned in the introduction and
earlier in this section, for OGLE-TR-122 and OGLE-TR-123 the authors
had to use stellar models to estimate the masses and radii of the
primary rather than the assumption of synchronization or
pseudo-synchronization (OGLE-TR-122 has an eccentric orbit), as the
latter would have implied masses and radii inconsistent with the
spectroscopic observations. We have included in Figure \ref{fig7}
eight published M dwarfs in double-lined binary systems (see Section
\ref{intro}), for a direct comparison with our results.

From Figure \ref{fig7} it appears that the values for the mass and
radius of T-Boo0-00080-B and T-Lyr1-01662-B are formally inconsistent
with theoretical predictions from the Lyon group
\citep{1998A&A...337..403B}. A marginal inconsistency is also observed
for T-Lyr0-08070-B, T-Cyg1-01385-B and T-Aur-013378-B. Our M dwarfs
have radii that are larger than predicted for the measured masses when
adopting consistent metallicities and ages. For the system
T-Boo0-00080, with a best fit metallicity of $\left[{\rm Fe/H}
\right]=+0.5$, a proper comparison is not possible because of the lack
of published isochrones for this composition. However, the solution
(based on synchronization) for solar metallicity is practically the
same as the one adopted, and in that case there is a clear
inconsistency. This radius discrepancy has been documented now for a
number of systems and in particular for several well-measured low-mass
double-lined eclipsing binaries (also included in Figure \ref{fig7}),
and is believed to reflect the fact that these systems do not evolve
as isolated stars. The rapid rotation, caused by tidal
synchronization, may lead to enhanced magnetic activity, which can
manifest itself in two ways: a decrease in the efficiency of energy
transport in these mostly convective stars, resulting in inflated
stellar radii and cooler temperatures, and significant spot coverage,
with similar consequences \citep[see, e.g.,][]{2005ApJ...631.1120L,
2006Ap&SS.304...89R, 2007ApJ...660..732L, 2007A&A...472L..17C}. In
order to check for possible activity, we looked for corresponding
X-ray sources in the ROSAT mission catalog
\citep{1999A&A...349..389V}, but no match was found for any of our
targets.

The next step in this work is to collect enough photometric data to
perform a similar analysis of the remaining 21 SEBs mentioned
earlier. Most of th observed eccentricities are close to zero (Figure
\ref{fig1}) and the distribution of rotational velocities vs. orbital
period suggest that orbit-rotation synchronization should be taking
place for many of these systems (see Figure \ref{fig2}). It would
strengthen the results to have detailed abundances for all these
targets in order to break the metallicity-temperature degeneracy that
affects our analysis in its present form. The recently commissioned
TRES instrument at FLWO \citep{2007RMxAC..28..129S,
2008ApJ...687.1253D} could provide suitable spectra for this
purpose. Absolute luminosities derived from accurate parallaxes would
significantly improve the determination of the mass and radius of the
primary stars during the isochrone modeling. The best hope for good
parallaxes is the GAIA mission \citep[and references
thereafter]{1994hpes.book.....B}, to be launched in 2011. With the
extra observations mentioned above, it should be possible to
unambiguously determine which SEBs are synchronized and which ones are
not, providing a valuable set of results for testing tidal theory.

\section{Summary}

We have determined masses and radii for the components of five
eclipsing single-lined binaries consisting of an F star and an unseen
M dwarf, identified photometrically by the TrES wide-angle transiting
planet survey. Our results are based on accurate spectroscopic orbital
solutions and high precision light curves, and were obtained in two
different ways: by modeling the primary star using stellar isochrones,
and by estimating the size of the primary star using the measured
value of $V_{\rm rot} \sin i_{\rm rot}$ together with the assumption
of synchronization and alignment between the spin axis of the primary
star and the orbital axis.

The near zero eccentricity of the orbits of these systems makes the
assumption of synchronization reasonable, following the predictions
from tidal theory. The consistency of the two sets of results depends
strongly on the adopted atmospheric parameters of the primary stars
during the modeling of their masses, in particular the
metallicity. Even when four of the five systems show an acceptable
match between the two sets of results, a definitive value for their
masses and radii still depends on the accurate determination of their
abundances. If we adopt the synchronization-based solutions that best
match those obtained from stellar evolution models, we find that in
four of the studied systems the results are inconsistent with low-mass
stellar evolution models, with M dwarfs that are larger than predicted
(two only marginally). This behavior has been documented previously
for a number of M dwarfs in binary systems.

Our results, combined with indications from previous work, show that
reliance on the assumption of synchronization to derive the mass and
radius of stars in eclipsing single-lined F+M is a useful tool, but
may not always be warranted and should be carefully tested against
stellar evolution models.

\acknowledgments{We thank Joe Zajac, Perry Berlind, and Mike Calkins
for obtaining some of the spectroscopic observations; Bob Davis for
maintaining the database for the CfA Digital Speedometers; and John
Geary, Andy Szentgyorgyi, Emilio Falco, Ted Groner, and Wayne Peters
for their contribution to making KeplerCam such an effective
instrument for obtaining high-quality light curves. We are also
grateful to the anonymous referee for very helpful comments and
suggestions. This research was supported in part by the Kepler Mission
under NASA Cooperative Agreement NCC2-1390. GT acknowledges partial
support from NSF grant AST-0708229}.

\bibliography{adssample}

\clearpage


\begin{deluxetable}{lllllll}
\tabletypesize{\small}
\tablewidth{0pt}
\tablecaption{\sc Target Coordinates and Magnitudes \label{coord_mag}}
\tablehead{
\colhead{} &
\colhead{} &
\colhead{T-Aur0-13378}&
\colhead{T-Boo0-00080}&
\colhead{T-Lyr1-01662}&
\colhead{T-Lyr0-08070}&
\colhead{T-Cyg1-01385}}
\startdata
RA  (2000)              &            & $ 05:05:06.9          $ & $ 14:35:54.5          $ & $ 18:59:02.8          $ & $ 19:19:03.7          $ & $ 20:15:21.9          $ \\
DEC (2000)              &            & $ +41:26:03           $ & $ +46:35:36           $ & $ +48:36:35           $ & $ +38:40:57           $ & $ +48:17:14           $ \\
$V$ (app. mag)          &            & $ 13.0                $ & $ 10.3                $ & $ 11.3                $ & $ 12.3                $ & $ 10.7                $ \\
$J-K_{\rm 2MASS}$ (mag) &            & $ 0.34                $ & $ 0.26                $ & $ 0.20                $ & $ 0.25                $ & $ 0.32                $ \\
\enddata
\end{deluxetable}

\begin{deluxetable}{crr}
\tablewidth{0pc}
\tablecaption{\sc Individual Radial Velocities for T-Aur0-13378 \label{rv_T-Aur0-13378}}
\tablehead{
\colhead{HJD}                    &
\colhead{$V_{\rm rad}$}          &
\colhead{$\sigma (V_{\rm rad})$} \\
\colhead{(days)}                 &
\colhead{(\kms)}                 &
\colhead{(\kms)}
}
\startdata
2453334.8407 &   $-$2.91  & 1.59 \\
2453401.7953 &     16.80  & 2.25 \\
2453452.6211 &  $-$18.19  & 2.00 \\
2453626.9868 &     24.28  & 2.04 \\
2453663.9614 &      9.94  & 1.57 \\
2453667.0023 &     36.17  & 1.37 \\
2453667.9707 &  $-$12.89  & 2.10 \\
2453684.9318 &     29.61  & 1.66 \\
2453687.9154 &     43.39  & 1.65 \\
2453721.9235 &  $-$10.47  & 1.66 \\
2453990.0023 &      1.47  & 1.66 \\
2454041.8878 &     40.07  & 1.90 \\
2454101.7741 &     32.28  & 3.05 \\
\enddata
\end{deluxetable}

\begin{deluxetable}{crr}
\tablewidth{0pc}
\tablecaption{\sc Individual Radial Velocities for T-Boo0-00080 \label{rv_T-Boo0-00080}}
\tablehead{
\colhead{HJD}                    &
\colhead{$V_{\rm rad}$}          &
\colhead{$\sigma (V_{\rm rad})$} \\
\colhead{(days)}                 &
\colhead{(\kms)}                 &
\colhead{(\kms)}
}
\startdata
2453078.7253 &  $-$41.87 & 1.11 \\ 
2453080.6658 &      1.13 & 0.78 \\
2453080.9194 &  $-$16.58 & 0.88 \\
2453084.6469 &  $-$15.00 & 1.16 \\
2453084.9011 &      3.86 & 1.09 \\
2453086.6660 &  $-$43.42 & 1.30 \\
2453086.8937 &  $-$35.05 & 1.90 \\
2453087.6528 &     16.50 & 1.05 \\
2453088.8989 &  $-$41.79 & 1.58 \\
2453094.6495 &  $-$26.89 & 1.90 \\
2453154.7784 &  $-$34.38 & 1.27 \\
2453176.6766 &     20.67 & 1.51 \\
2453187.7755 &  $-$32.58 & 1.16 \\
2453227.5743 &     19.77 & 1.09 \\
2453780.0387 &  $-$45.40 & 1.62 \\
2453807.9210 &  $-$45.37 & 1.54 \\
2453808.8949 &     11.24 & 1.13 \\
2453810.8555 &  $-$26.81 & 1.14 \\
\enddata
\end{deluxetable}

\begin{deluxetable}{crr}
\tablewidth{0pc}
\tablecaption{\sc Individual Radial Velocities for T-Lyr1-01662 \label{rv_T-Lyr1-01662}}
\tablehead{
\colhead{HJD}                    &
\colhead{$V_{\rm rad}$}          &
\colhead{$\sigma (V_{\rm rad})$} \\
\colhead{(days)}                 &
\colhead{(\kms)}                 &
\colhead{(\kms)}
}
\startdata
2453661.7019 &  $-$28.11 & 0.47 \\
2453663.6613 &     17.34 & 0.69 \\
2453666.6132 &   $-$6.53 & 0.82 \\
2453684.5817 &     18.79 & 0.81 \\
2453688.5651 &     15.52 & 0.95 \\
2453691.5797 &  $-$21.79 & 1.00 \\
2453694.5962 &  $-$26.56 & 0.97 \\
2453837.9580 &   $-$2.49 & 0.92 \\
2453838.9680 &  $-$33.90 & 1.01 \\
2453840.9543 &     16.64 & 0.92 \\
2453842.9740 &  $-$30.62 & 1.00 \\
2453866.9636 &     15.00 & 0.84 \\
2453871.9095 &   $-$6.71 & 1.19 \\
2453987.6572 &  $-$29.91 & 0.93 \\
2453992.6429 &   $-$3.20 & 0.85 \\
\enddata
\end{deluxetable}

\begin{deluxetable}{crr}
\tablewidth{0pc}
\tablecaption{\sc Individual Radial Velocities for T-Lyr0-08070 \label{rv_T-Lyr0-08070}}
\tablehead{
\colhead{HJD}                    &
\colhead{$V_{\rm rad}$}          &
\colhead{$\sigma (V_{\rm rad})$} \\
\colhead{(days)}                 &
\colhead{(\kms)}                 &
\colhead{(\kms)}
}
\startdata
2453129.9811 &  $-$47.72 & 3.95 \\
2453135.9131 &  $-$45.10 & 4.41 \\
2453157.9641 &  $-$39.31 & 2.90 \\
2453160.8934 &  $-$26.61 & 2.05 \\
2453457.0047 &  $-$47.04 & 3.93 \\
2453481.9792 &  $-$23.26 & 3.01 \\
2453483.9629 &  $-$61.90 & 4.02 \\
2453485.9666 &      4.96 & 2.71 \\
2453486.9020 &     13.84 & 2.42 \\
2453487.9680 &   $-$6.14 & 2.98 \\
2453488.9413 &  $-$48.73 & 2.76 \\
2453509.9377 &  $-$55.56 & 3.90 \\
2453510.8796 &   $-$1.92 & 2.58 \\
2453511.9144 &     13.61 & 2.18 \\
2453872.9397 &  $-$30.64 & 6.23 \\
\enddata
\end{deluxetable}

\begin{deluxetable}{crr}
\tablewidth{0pc}
\tablecaption{\sc Individual Radial Velocities for T-Cyg1-01385 \label{rv_T-Cyg1-01385}}
\tablehead{
\colhead{HJD}                    &
\colhead{$V_{\rm rad}$}          &
\colhead{$\sigma (V_{\rm rad})$} \\
\colhead{(days)}                 &
\colhead{(\kms)}                 &
\colhead{(\kms)}
}
\startdata
2453333.6202 &      8.44 & 1.17 \\
2453508.9739 &  $-$38.15 & 1.19 \\
2453511.9801 &     24.75 & 0.91 \\
2453540.9615 &  $-$38.52 & 1.72 \\
2453548.9596 &  $-$24.14 & 1.33 \\
2453576.7477 &     18.30 & 1.13 \\
2453626.7511 &  $-$41.05 & 1.33 \\
2453629.6988 &     22.52 & 1.00 \\
2453632.7413 &  $-$36.62 & 1.64 \\
2453657.6540 &   $-$2.19 & 1.12 \\
2453658.7019 &  $-$30.46 & 1.10 \\
2453659.6389 &  $-$40.07 & 1.45 \\
2453660.6781 &  $-$20.03 & 1.40 \\
2453661.7327 &     11.68 & 1.57 \\
\enddata
\end{deluxetable}

\begin{deluxetable}{lllllll}
\tabletypesize{\footnotesize}
\tablewidth{0pt}
\tablecaption{\sc Spectroscopic Orbital Solutions \label{rv_results}}
\tablehead{
\colhead{} &
\colhead{} &
\colhead{T-Aur0-13378}&
\colhead{T-Boo0-00080}&
\colhead{T-Lyr1-01662}&
\colhead{T-Lyr0-08070}&
\colhead{T-Cyg1-01385}}
\startdata
N observations        &            & $ 14                  $ & $ 18                  $ & $ 13                  $ & $ 16                  $ & $ 15                  $ \\
Time span             & days       & $ 802                 $ & $ 732                 $ & $ 380                 $ & $ 863                 $ & $ 656                 $ \\
$P$                   & days       & $ 3.54182(14)         $ & $ 2.539825(51)        $ & $ 4.23339(22)         $ & $ 1.184780(25)        $ & $ 6.56012(29)         $ \\
$\gamma$              & \kms       & $ +11.47   \pm 0.44   $ & $ -12.97  \pm 0.30    $ & $ -7.42    \pm 0.19   $ & $ -25.49    \pm 0.98  $ & $ -6.81     \pm 0.17  $ \\
$K$                   & \kms       & $ 32.31    \pm 0.70   $ & $ 32.81   \pm 0.38    $ & $ 26.44    \pm 0.26   $ & $ 42.87     \pm 1.50  $ & $ 33.62     \pm 0.21  $ \\
$e$                   &            & $ 0.040    \pm 0.022  $ & $ 0.026   \pm 0.012   $ & $ 0.037    \pm 0.010  $ & $ 0.054     \pm 0.036 $ & $ 0.023     \pm 0.007 $ \\
$\omega$              & $^{\circ}$ & $149       \pm 25     $ & $ 78      \pm 25      $ & $ 133      \pm 16     $ & $ 346       \pm 74    $ & $ 159       \pm 17    $ \\    
$O-C$ RMS             & \kms       & $ 1.47                $ & $ 1.06                $ & $ 0.73                $ & $ 3.38                $ & $ 0.54                $ \\
\enddata
\end{deluxetable}

\begin{deluxetable}{lllllll}
\tabletypesize{\footnotesize}
\tablewidth{0pt}
\tablecaption{\sc KeplerCam Photometry \label{kepcam_phot}}
\tablehead{
\colhead{} &
\colhead{} &
\colhead{T-Aur0-13378}&
\colhead{T-Boo0-00080}&
\colhead{T-Lyr1-01662}&
\colhead{T-Lyr0-08070}&
\colhead{T-Cyg1-01385}}
\startdata
Date (UT)             &            & 2006 Dec 12             &  2006 Apr 17            & 2006 Jul 06             & 2006 Sep 18             & 2006 Jul 10             \\
Band                  &            & Sloan $i$               &  Sloan $i$              & Sloan $i$               & Sloan $z$               & Sloan $z$               \\
Exposure              & sec        & $ 30                  $ & $ 30                  $ & $ 30                  $ & $ 45                  $ & $ 30                  $ \\
Cadence               & im/min     & $ 1.42                $ & $ 1.42                $ & $ 1.42                $ & $ 1.04                $ & $ 1.42                $ \\
FWHM                  & arcsec     & $ 1.4 - 1.9           $ & $ 1.7 - 2.6           $ & $ 1.4 - 1.9           $ & $ 1.6 - 2.1           $ & $ 1.4 - 2.7           $ \\
$\sec z$              &            & $ 1.0 - 2.3           $ & $ 1.7 - 1.0 - 1.2     $ & $ 1.4 - 1.1 - 1.2     $ & $ 1.0 - 1.7           $ & $ 1.6 - 1.0 - 1.3     $ \\
\enddata
\end{deluxetable}

\begin{deluxetable}{ccccc}
\tabletypesize{\footnotesize}
\tablewidth{0pc}
\tablecaption{\sc KeplerCam Light Curves \label{kepcam_lc}}
\tablehead{
\colhead{Object}                 &
\colhead{Band}                   &
\colhead{HJD}                    &
\colhead{Relative Flux}          &
\colhead{$\sigma$}
}
\startdata
 T-Aur0-13378 & i & 2454081.768872 & 0.9982 & 0.0016 \\
 T-Aur0-13378 & i & 2454081.769381 & 0.9989 & 0.0016 \\
 T-Aur0-13378 & i & 2454081.769890 & 0.9954 & 0.0016 \\
 T-Aur0-13378 & i & 2454081.770400 & 0.9942 & 0.0016 \\
  & & & & \\
 T-Boo0-00080 & i & 2453842.652138 & 0.9998 & 0.0014 \\
 T-Boo0-00080 & i & 2453842.652844 & 0.9986 & 0.0014 \\
 T-Boo0-00080 & i & 2453842.653319 & 0.9974 & 0.0014 \\
 T-Boo0-00080 & i & 2453842.653805 & 0.9996 & 0.0014 \\
 & & & & \\
 T-Lyr1-01662 & i & 2453922.652243 & 1.0001 & 0.0015 \\
 T-Lyr1-01662 & i & 2453922.652729 & 0.9947 & 0.0015 \\
 T-Lyr1-01662 & i & 2453922.653227 & 0.9967 & 0.0015 \\
 T-Lyr1-01662 & i & 2453922.653690 & 1.0003 & 0.0015 \\
 & & & & \\
 T-Lyr0-08070 & z & 2453996.632634 & 0.9984 & 0.0015 \\
 T-Lyr0-08070 & z & 2453996.633282 & 0.9982 & 0.0015 \\
 T-Lyr0-08070 & z & 2453996.633954 & 0.9982 & 0.0015 \\
 T-Lyr0-08070 & z & 2453996.634602 & 1.0014 & 0.0015 \\
 & & & & \\
 T-Cyg1-01385 & z & 2453926.673722 & 0.9976 & 0.0012 \\
 T-Cyg1-01385 & z & 2453926.674324 & 1.0021 & 0.0012 \\
 T-Cyg1-01385 & z & 2453926.675527 & 0.9966 & 0.0012 \\
 T-Cyg1-01385 & z & 2453926.676129 & 1.0016 & 0.0012 \\
\enddata
\tablecomments{
These are sample entries of the full light curves. The complete versions are given on-line.}
\end{deluxetable}

\begin{deluxetable}{lllllll}
\tabletypesize{\footnotesize}
\tablewidth{0pt}
\tablecaption{\sc Primary Star Spectroscopic Atmospheric Properties \label{atm_prop}}
\tablehead{
\colhead{} &
\colhead{} &
\colhead{T-Aur0-13378}&
\colhead{T-Boo0-00080}&
\colhead{T-Lyr1-01662}&
\colhead{T-Lyr0-08070}&
\colhead{T-Cyg1-01385}}
\startdata
\cutinhead{$[\rm Fe/H] \ _{\rm adopted} \ = \  -1.0$}
$T_{\rm eff}$              & K            & $ 5860 \pm 100        $ & $ 5510 \pm 30            $ & $ 5810 \pm 30           $ & $ 6000 \pm 150        $ & $ 5520 \pm 30           $ \\
$V_{\rm rot}\sin i_{\rm rot}$ & \kms      & $ 25.3  \pm 0.7       $ & $ 35.9 \pm 0.5           $ & $ 13.0 \pm 0.4          $ & $ 57.9 \pm 2.1        $ & $ 11.9 \pm 0.6          $ \\
${\rm log} \ g_{\rm A}$  & \cmss          & $ 3.79 \pm 0.02       $ & $ 4.01 \pm 0.02 $ \tablenotemark{a} & $ 4.22 \pm 0.04         $ & $ 4.13 \pm 0.01       $ & $ 3.98 \pm 0.02 $ \tablenotemark{a} \\
\cutinhead{$[\rm Fe/H] \ _{\rm adopted} \ = \  -0.5$}
$T_{\rm eff}$              & K            & $ 6200 \pm 80         $ & $ 5850 \pm 30            $ & $ 6200 \pm 30           $ & $ 6250 \pm 140        $ & $ 5580 \pm 30           $ \\
$V_{\rm rot}\sin i_{\rm rot}$ & \kms      & $ 25.7  \pm 0.7       $ & $ 35.8 \pm 0.5           $ & $ 13.6 \pm 0.4          $ & $ 57.9 \pm 2.1        $ & $ 12.6 \pm 0.6          $ \\
${\rm log} \ g_{\rm A}$  & \cmss          & $ 3.83 \pm 0.02       $ & $ 4.01 \pm 0.02          $ & $ 4.25 \pm 0.03         $ & $ 4.16 \pm 0.01       $ & $ 3.98 \pm 0.02         $ \\
\cutinhead{$[\rm Fe/H] \ _{\rm adopted} \ = \  0.0$}
$T_{\rm eff}$              & K            & $ 6620 \pm 80         $ & $ 6190 \pm 30            $ & $ 6760 \pm 30           $ & $ 6500 \pm 140        $ & $ 6270 \pm 30           $ \\
$V_{\rm rot}\sin i_{\rm rot}$ & \kms      & $ 25.5  \pm 0.7       $ & $ 35.8 \pm 0.5           $ & $ 13.8 \pm 0.4          $ & $ 57.8 \pm 2.1        $ & $ 13.0 \pm 0.6          $ \\
${\rm log} \ g_{\rm A}$  & \cmss          & $ 3.88 \pm 0.02       $ & $ 4.06 \pm 0.02          $ & $ 4.30 \pm 0.02         $ & $ 4.21 \pm 0.01       $ & $ 4.03 \pm 0.02         $ \\
\cutinhead{$[\rm Fe/H] \ _{\rm adopted} \ = \  +0.5$}
$T_{\rm eff}$              & K            & $ 6990 \pm 80         $ & $ 6610 \pm 30            $ & $ 7210 \pm 40           $ & $ 6710 \pm 140        $ & $ 6710 \pm 30           $ \\
$V_{\rm rot}\sin i_{\rm rot}$ & \kms      & $ 25.5  \pm 0.7       $ & $ 35.9 \pm 0.5           $ & $ 14.4 \pm 0.4          $ & $ 58.3 \pm 2.1        $ & $ 13.4 \pm 0.6          $ \\
${\rm log} \ g_{\rm A}$  & \cmss          & $ 3.91 \pm 0.02       $ & $ 4.10 \pm 0.02          $ & $ 4.30 \pm 0.02 $ \tablenotemark{a} & $ 4.23 \pm 0.01       $ & $ 4.07 \pm 0.01         $ \\
\enddata
\tablenotetext{a}{
Surface gravity adopted from isochrone fits corresponding to the
nearest metallicity that gives a meaningful result}.
\end{deluxetable}

\begin{deluxetable}{lllllll}
\tabletypesize{\footnotesize}
\tablewidth{0pt}
\tablecaption{\sc Summary of Parameter of Transit Light Curve Analysis \label{lc_fit}}
\tablehead{
\colhead{} &
\colhead{} &
\colhead{T-Aur0-13378}&
\colhead{T-Boo0-00080}&
\colhead{T-Lyr1-01662}&
\colhead{T-Lyr0-08070}&
\colhead{T-Cyg1-01385}}
\startdata
Epoch                 & HJD        & $ 2454081.8678(2)     $ & $ 2453842.8212(2)     $ & $ 2453922.7207(2)     $ & $ 2453996.7368(2)     $ & $ 2453926.8104(2)     $ \\ 
Duration              & min        & $ 156.9  \pm 0.4      $ & $ 194.1  \pm 0.2      $ & $ 201.4  \pm 0.3      $ & $ 176.6  \pm 0.2      $ & $ 190.2  \pm 0.2      $ \\ 
$u_{1}$                   &        & $ 0.1509              $ & $ 0.1917              $ & $ 0.1628              $ & $ 0.2467              $ & $ 0.1431              $ \\ 
$u_{2} $                  &        & $ 0.3834              $ & $ 0.3677              $ & $ 0.3776              $ & $ 0.3821              $ & $ 0.3594              $ \\
$a/R_{\rm A}$         &            & $ 5.10 \pm 0.08       $ & $ 5.21 \pm 0.03       $ & $ 9.56 \pm 0.08       $ & $ 3.69 \pm 0.01       $ & $ 9.76^{+0.05}_{-0.08} $ \\
$R_{\rm B}/R_{\rm A}$ &            & $ 0.1551(7)           $ & $ 0.1775(8)           $ & $ 0.2085(8)           $ & $ 0.1952(4)           $ & $ 0.2205(3)           $ \\
$b$                   &            & $ 0.63  \pm 0.02      $ & $ 0.815  \pm 0.004    $ & $ 0.716  \pm 0.007    $ & $ 0.05^{+0.05}_{-0.03}$ & $ 0.11  \pm 0.06      $ \\
$O-C$ RMS             & \%         & $ 0.21                $ & $ 0.10                $ & $ 0.16                $ & $ 0.18                $ & $ 0.22                $ \\
\hline \\
$\rho_{\rm A} $ \tablenotemark{a} & $\rm gr \ cm^{-3}$   & $ 0.163 \pm 0.008     $ & $ 0.344 \pm 0.006     $ & $ 0.75 \pm 0.02       $ & $ 0.542 \pm 0.008     $ & $  0.297 \pm 0.007    $ \\
${\rm log} \ g_{\rm B} $ \tablenotemark{a} & \cmss    & $ 4.86 \pm 0.02       $ & $ 4.91 \pm 0.01       $ & $ 4.98 \pm 0.01       $ & $ 4.97 \pm 0.02       $ & $  4.86 \pm 0.01      $ \\
\enddata 
\tablenotetext{a}{ 
These quantities are essentially model-independent and rely only
on spectroscopic and photometric observables. The dependence of
$\rho_{\rm A}$ on $\rho_{\rm B}$ is typically very weak as the
secondary star is usually small compared to the primary (See
Eq. \ref{density}).}
\end{deluxetable}

\begin{deluxetable}{lllllll}
\tabletypesize{\footnotesize}
\tablewidth{0pt}
\tablecaption{\sc Results derived from Stellar Isochrones \label{iso_results}}
\tablehead{
\colhead{} &
\colhead{} &
\colhead{T-Aur0-13378}&
\colhead{T-Boo0-00080 \tablenotemark{a}}&
\colhead{T-Lyr1-01662 \tablenotemark{a}}&
\colhead{T-Lyr0-08070}&
\colhead{T-Cyg1-01385 \tablenotemark{a}}}
\startdata
\cutinhead{$[\rm Fe/H] \ _{\rm adopted} \ = \  -1.0 \pm 0.2 $}
Age                     & Gyr       & $ 9.0   \pm 3.0         $ & $ -                   $ & $ 12.9 \pm 1.5          $ & $ 12.6 \pm 1.8        $ & $ -                   $ \\
$M_{\rm A}$             & \msun     & $ 0.92  \pm 0.11        $ & $ -                   $ & $ 0.81 \pm 0.20         $ & $ 0.82 \pm 0.06       $ & $ -                   $ \\
$R_{\rm A}$             & \rsun     & $ 2.03  \pm 0.08        $ & $ -                   $ & $ 1.16 \pm 0.08         $ & $ 1.30 \pm 0.03       $ & $ -                   $ \\
$M_{\rm B}$             & \msun     & $ 0.26  \pm 0.02        $ & $ -                   $ & $ 0.20 \pm 0.03         $ & $ 0.22 \pm 0.02       $ & $ -                   $ \\
$R_{\rm B}$             & \rsun     & $ 0.31  \pm 0.02        $ & $ -                   $ & $ 0.24 \pm 0.02         $ & $ 0.25 \pm 0.01       $ & $ -                   $ \\
\cutinhead{$[\rm Fe/H] \ _{\rm adopted} \ = \  -0.5 \pm 0.2 $}
Age                     & Gyr       & $ 4.5   \pm 1.3         $ & $ 10.4 \pm 3.7        $ & $ 7.7  \pm 4.8          $ & $ 6.9  \pm 1.7        $ & $ 9.5  \pm 2.6        $ \\
$M_{\rm A}$             & \msun     & $ 1.18  \pm 0.14        $ & $ 0.93 \pm 0.15       $ & $ 0.96 \pm 0.18         $ & $ 1.00 \pm 0.08       $ & $ 0.95 \pm 0.10       $ \\
$R_{\rm A}$             & \rsun     & $ 2.18  \pm 0.08        $ & $ 1.58 \pm 0.07       $ & $ 1.22 \pm 0.07         $ & $ 1.38 \pm 0.03       $ & $ 1.66 \pm 0.05       $ \\
$M_{\rm B}$             & \msun     & $ 0.30  \pm 0.02        $ & $ 0.24 \pm 0.02       $ & $ 0.23 \pm 0.03         $ & $ 0.25 \pm 0.02       $ & $ 0.35 \pm 0.02       $ \\
$R_{\rm B}$             & \rsun     & $ 0.34  \pm 0.02        $ & $ 0.28 \pm 0.02       $ & $ 0.25 \pm 0.02         $ & $ 0.27 \pm 0.01       $ & $ 0.37 \pm 0.01       $ \\
\cutinhead{$[\rm Fe/H] \ _{\rm adopted} \ = \  0.0 \pm 0.2 $}
Age                     & Gyr       & $ 2.0   \pm 1.5         $ & $ 4.1 \pm 1.7         $ & $ 0.8  \pm 0.7          $ & $ 2.2  \pm 0.7        $ & $ 3.7  \pm 1.5        $ \\
$M_{\rm A}$             & \msun     & $ 1.60  \pm 0.13        $ & $ 1.26 \pm 0.17       $ & $ 1.35 \pm 0.08         $ & $ 1.33 \pm 0.07       $ & $ 1.31 \pm 0.16       $ \\
$R_{\rm A}$             & \rsun     & $ 2.40  \pm 0.10        $ & $ 1.74 \pm 0.07       $ & $ 1.35 \pm 0.04         $ & $ 1.50 \pm 0.03       $ & $ 1.82 \pm 0.06       $ \\
$M_{\rm B}$             & \msun     & $ 0.37  \pm 0.03        $ & $ 0.28 \pm 0.02       $ & $ 0.28 \pm 0.02         $ & $ 0.29 \pm 0.02       $ & $ 0.43 \pm 0.03       $ \\
$R_{\rm B}$             & \rsun     & $ 0.37  \pm 0.02        $ & $ 0.31 \pm 0.02       $ & $ 0.28 \pm 0.01         $ & $ 0.29 \pm 0.01       $ & $ 0.40 \pm 0.02       $ \\
\cutinhead{$[\rm Fe/H] \ _{\rm adopted} \ = \  +0.5 \pm 0.2 $}
Age                     & Gyr       & $ 0.9   \pm 0.2         $ & $ 1.2 \pm 0.5         $ & $ -                     $ & $ 0.5  \pm 0.1        $ & $ 1.0  \pm 0.4        $ \\
$M_{\rm A}$             & \msun     & $ 1.91  \pm 0.07        $ & $ 1.58 \pm 0.06       $ & $ -                     $ & $ 1.52 \pm 0.02       $ & $ 1.64 \pm 0.06       $ \\
$R_{\rm A}$             & \rsun     & $ 2.54  \pm 0.05        $ & $ 1.86 \pm 0.02       $ & $ -                     $ & $ 1.57 \pm 0.01       $ & $ 1.95 \pm 0.03       $ \\
$M_{\rm B}$             & \msun     & $ 0.41  \pm 0.01        $ & $ 0.33 \pm 0.01       $ & $ -                     $ & $ 0.32 \pm 0.01       $ & $ 0.49 \pm 0.01       $ \\
$R_{\rm B}$             & \rsun     & $ 0.39  \pm 0.01        $ & $ 0.33 \pm 0.01       $ & $ -                     $ & $ 0.31 \pm 0.01       $ & $ 0.43 \pm 0.01       $ \\
\enddata 
\tablenotetext{a}{
No meaningful solution found for T-Boo0-00080, T-Lyr1-01662 and
T-Cyg1-01385 when adopting a very low or high metallicity for the
primary star.}
\end{deluxetable}

\begin{deluxetable}{lllllll}
\tabletypesize{\footnotesize}
\tablewidth{0pt}
\tablecaption{\sc Results derived from Orbit-Rotation Synchronization \label{sync_results}}
\tablehead{
\colhead{} &
\colhead{} &
\colhead{T-Aur0-13378 \tablenotemark{a}}&
\colhead{T-Boo0-00080}&
\colhead{T-Lyr1-01662}&
\colhead{T-Lyr0-08070}&
\colhead{T-Cyg1-01385}}
\startdata
$[\rm Fe/H] \ _{\rm adopted} $ \tablenotemark{b} &                 & $ -                   $ & $ +0.5                $ & $ -0.5                $ & $ -0.5                $ & $ -0.5                $ \\
$M_{\rm A (sync)}$             & \msun                & $ -                   $ & $ 1.49  \pm 0.07      $ & $ 0.77 \pm 0.08       $ & $ 0.95 \pm 0.11       $ & $ 0.91  \pm 0.15      $ \\
$R_{\rm A (sync)}$             & \rsun                & $ -                   $ & $ 1.83  \pm 0.03      $ & $ 1.14 \pm 0.03       $ & $ 1.36 \pm 0.05       $ & $ 1.63  \pm 0.08      $ \\
${\rm log} \ g_{\rm A (sync)}$ & \cmss                & $ -                   $ & $ 4.09  \pm 0.01      $ & $ 4.21 \pm 0.02       $ & $ 4.15 \pm 0.02       $ & $ 3.97  \pm 0.03      $ \\
$M_{\rm B (sync)}$             & \msun                & $ -                   $ & $ 0.315  \pm 0.010    $ & $ 0.198 \pm 0.012     $ & $ 0.240 \pm 0.019     $ & $ 0.345  \pm 0.034    $ \\
$R_{\rm B (sync)}$             & \rsun                & $ -                   $ & $ 0.325  \pm 0.005    $ & $ 0.238 \pm 0.007     $ & $ 0.265 \pm 0.010     $ & $ 0.360  \pm 0.017    $ \\
\enddata \tablenotetext{a}{ 
No match found between solutions based on synchronization and stellar
models for the system T-Aur0-13378.}  
\tablenotetext{b}{ 
Metallicity for the best match between the solutions based on
synchronization and stellar models.}
\end{deluxetable}

\clearpage


\clearpage

\clearpage

\end{document}